\documentclass[twocolumn]{aastex631}

\usepackage{CJKutf8}
\usepackage{amsmath}
\usepackage[inline]{enumitem}

\received{January 17, 2025}
\revised{February 6, 2025}
\accepted{May 15, 2025}

\submitjournal{ApJ}

\shorttitle{Identifying CT-AGNs with ML algorithm in CDFS}
\shortauthors{Zhang et al.}

\begin{document}

\title{Identifying Compton-thick AGNs with Machine learning algorithm in \textit{Chandra} Deep Field-South}

\correspondingauthor{Xiaotong Guo; Qiusheng Gu}
\email{guoxiaotong@aqnu.edu.cn; \\qsgu@nju.edu.cn}

\author{Rui Zhang \begin{CJK*}{UTF8}{gkai}(张蕊)\end{CJK*}}
\affiliation{School of Mathematics and Physics, Anqing Normal University, Anqing 246133, China}
\affiliation{Institute of Astronomy and Astrophysics, Anqing Normal University, Anqing 246133, China}

\author[0000-0002-2338-7709]{Xiaotong Guo \begin{CJK*}{UTF8}{gkai}(郭晓通)\end{CJK*}}
\affiliation{School of Mathematics and Physics, Anqing Normal University, Anqing 246133, China}
\affiliation{Institute of Astronomy and Astrophysics, Anqing Normal University, Anqing 246133, China}

\author[0000-0002-3890-3729]{Qiusheng Gu \begin{CJK*}{UTF8}{gkai}(顾秋生)\end{CJK*}}
\affiliation{School of Astronomy and Space Science, Nanjing University, Nanjing, Jiangsu 210093, China}
\affiliation{Key Laboratory of Modern Astronomy and Astrophysics (Nanjing University), Ministry of Education, Nanjing 210093, China}

\author[0000-0001-9694-2171]{Guanwen Fang \begin{CJK*}{UTF8}{gkai}(方官文)\end{CJK*}}
\affiliation{School of Mathematics and Physics, Anqing Normal University, Anqing 246133, China}
\affiliation{Institute of Astronomy and Astrophysics, Anqing Normal University, Anqing 246133, China}

\author[0000-0003-1697-6801]{Jun Xu \begin{CJK*}{UTF8}{gkai}(徐骏)\end{CJK*}}
\affiliation{School of Mathematics and Physics, Anqing Normal University, Anqing 246133, China}
\affiliation{Institute of Astronomy and Astrophysics, Anqing Normal University, Anqing 246133, China}

\author[0000-0002-1530-2680]{Hai-Cheng Feng} 
\affiliation{Yunnan Observatories, Chinese Academy of Sciences, Kunming 650216, China}

\author[0000-0001-5895-0189]{Yongyun Chen \begin{CJK*}{UTF8}{gkai}(陈永云)\end{CJK*}}
\affiliation{College of Physics and Electronic Engineering, Qujing Normal University, Qujing 655011, China}

\author[0000-0002-3490-4089]{Rui Li}
\affiliation{Institude for Astrophysics, School of Physics, Zhengzhou University, Zhengzhou, 450001, China}

\author[0000-0003-1028-8733]{Nan Ding \begin{CJK*}{UTF8}{gkai}(丁楠)\end{CJK*}}
\affiliation{School of Physical Science and Technology, Kunming University, Kunming 650214, China}

\author[0000-0002-3023-8226]{Hongtao Wang \begin{CJK*}{UTF8}{gkai}(王洪涛)\end{CJK*}}
\affiliation{School of Science, Langfang Normal University,Langfang 065000, China}


%
%
%



\begin{abstract}
Compton-thick active galactic nuclei (CT-AGNs), which are defined by column density $\mathrm{N_H} \geqslant 1.5 \times 10^{24} \ \mathrm{cm}^{-2}$, emit feeble X-ray radiation, even undetectable by X-ray instruments. Despite this, the X-ray emissions from CT-AGNs are believed to be a substantial contributor to the cosmic X-ray background (CXB). According to synthesis models of AGNs, CT-AGNs are expected to make up a significant fraction of the AGN population, likely around 30\% or more. However, only $\sim$11\% of AGNs have been identified as CT-AGNs in the Chandra Deep Field-South (CDFS). To identify hitherto unknown CT-AGNs in the field, we used a Random Forest algorithm for identifying them. First, we build a secure classified subset of 210 AGNs to train and evaluate our algorithm. Our algorithm achieved an accuracy rate of 90\% on the test set after training. Then, we applied our algorithm to an additional subset of 254 AGNs, successfully identifying 67 CT-AGNs within this group. This result significantly increased the fraction of CT-AGNs in the CDFS, which is closer to the theoretical predictions of the CXB. Finally, we compared the properties of host galaxies between CT-AGNs and non-CT-AGNs and found that the host galaxies of CT-AGNs exhibit higher levels of star formation activity.


\end{abstract}



\section{Introduction} \label{sec:intro}

Active galactic nuclei (AGNs) are powered by the accretion of surrounding matter through supermassive black holes (SMBHs). The emissions from AGNs span across the entire electromagnetic spectrum, exhibiting distinct radiation characteristics at different distances from the central SMBHs. The AGN unification model posits that the radiation across various bands of an AGN originates from its distinct substructures \citep[e.g.,][]{1993ARA&A..31..473A,1995PASP..107..803U,2015ARA&A..53..365N}.
For example, the accretion disk near a SMBH heats up by releasing its gravitational potential energy, thereby emitting radiation across the ultraviolet (UV) to optical spectrum \citep[e.g.,][]{1980ApJ...235..361R}; the corona above or around the accretion disk is thought for the production of X-ray radiation through inverse Compton scattering \citep[e.g.,][]{1979ApJ...234.1105L,1991ApJ...380L..51H};
the dusty torus outside the corona and accretion disk absorbs the intense emissions from the center and then re-emits this energy at longer, MIR wavelengths \citep[e.g.,][]{1993ARA&A..31..473A}.
In addition, the classification of AGNs usually uses the observation features of various bands. For instance, based on the characteristics of the optical emission lines, AGNs can be classified into types 1 and 2 \citep[e.g.,][]{1974ApJ...192..581K}. Based on the column density ($\mathrm{N_H}$) of the X-ray absorption, AGNs can be classified into Compton-thin AGNs ($\mathrm{N_H} < 1.5\times 10^{24}\ \mathrm{cm}^{-2}$) and Compton-thick AGNs \citep[$\mathrm{N_H} \geqslant  1.5\times 10^{24}\ \mathrm{cm}^{-2}$, CT-AGNs; e.g.,][]{2004ASSL..308..245C}.

The luminous and persistent X-ray emission is an important observational feature that distinguishes AGNs from other extragalactic sources. Their X-ray radiations are recognized as the predominant contributors to the cosmic X-ray background (CXB), with up to 93\% of the extragalactic CXB at energies below 10 keV has been resolved into AGNs \citep[e.g.,][]{2003ApJ...588..696M, 2005MNRAS.357.1281W, 2023ExA....56..141L}. 
As a subclass of AGNs, CT-AGNs are naturally considered an important origin of the CXB. 
Due to heavy absorption in their soft X-ray band, their primary contribution to the CXB is around the peak of the CXB \citep[$\sim$ 30~keV;][]{2008ApJ...689..666A}, with approximately 10\%--25\% of the radiation being generated by CT-AGNs.
\cite{2011ApJ...728...58B}  used X-ray radiation from AGNs across different absorption levels to model the CXB, revealing that CT-AGNs account for $20^{+9}_{-6}\%$ of the intrinsic AGN population.
In some models, a higher fraction of CT-AGNs is required, even reaching 50\% of the total AGN population \citep[e.g.,][]{2014ApJ...786..104U, 2015ApJ...802...89B, 2019ApJ...871..240A}.
The observed fraction of CT-AGNs recovered within $z<0.01$ is $20\% \pm 5\%$ \citep{2021ApJ...922..252T}, which aligns with the predicted fraction from the AGN synthetic model of the CXB. 
In high-redshift environments, a higher fraction of CT-AGNs should be expected, where the mass fraction of atomic and/or molecular gas in galaxies is much higher than in the local universe \citep[e.g.,][]{2013ARA&A..51..105C}.
However, only a small fraction ($\lesssim$ 10\%) of CT-AGNs is identified in deep X-ray surveys \citep[e.g.,][]{2016ApJ...817...34M, 2018MNRAS.480.2578L, 2015A&A...573A.137L}. Even Nuclear Spectroscopic Telescope Array \citep[NuSTAR,][]{2013ApJ...770..103H}, which is sensitive above 10 keV, was only able to increase the fraction of CT-AGNs to $\sim 11.5\%$ in the UKIDSS Ultra Deep Survey field \citep{2018ApJS..235...17M}. Therefore, a large number of CT-AGNs remain unidentified or undiscovered in deep X-ray surveys.

X-ray spectroscopy fitting is the most commonly used method to  identify CT-AGNs. For instance, \cite{2018MNRAS.480.2578L} identified 67 CT-AGNs from 1855 AGNs in the \textit{Chandra}-COSMOS. 
Moreover, the combination of NuSTAR and Chandra or XMM-Newton observations will constitute a broad waveband X-ray spectrum, which can be better used to identify CT-AGNs.
Many recent studies have used the broad waveband X-ray spectroscopy fitting to identify CT-AGNs \citep[e.g.,][]{2019ApJ...877..102K, 2019MNRAS.487.1662L, 2019ApJ...887..173L, 2020ApJ...888....8T,2021ApJ...922..252T,2022ApJ...935..114M,2022ApJ...940..148S}. 
Several multiwavelength techniques are also used to identify CT-AGNs, which rely on the ratio between X-ray and other wavelength bands, including X-ray to MIR luminosity ratio \citep[e.g.,][]{2015A&A...578A.120L, 2015ApJ...809..115L, 2017ApJ...846...20L} and X-ray to high-ionization optical emission lines luminosity ratio \citep[e.g.,][]{1998A&A...338..781M, 2006A&A...446..459C, 2010A&A...519A..92G, 2015A&A...578A.120L}. 
In recent years, the rapid development of machine learning (ML) algorithms has brought revolutionary changes to various fields, including astronomy. ML algorithms are also used to identify CT-AGN. 
For example, \cite{2023A&A...675A..65S} used a ML algorithm for reliably predicting column densities, and tried to identify CT-AGNs.

The \textit{Chandra} Deep Field-South (CDFS) survey is currently the deepest X-ray survey with an exposure time of about 7~Ms \citep{2017ApJS..228....2L}. 
It is expected that the fraction of CT-AGNs identified in this survey will be consistent with the theoretical predictions for the CXB. There are 71 CT-AGNs that have been identified through X-ray spectroscopy fitting \citep{2017ApJS..232....8L, 2019ApJ...877....5L, 2019A&A...629A.133C}. 
However, the fraction of CT-AGNs identified in the CDFS survey still falls short of the theoretical predictions for the CXB.
\cite{2020ApJ...897..160L} pointed out that a large population of obscured AGNs was misdiagnosed as low-luminosity AGNs in the CDFS survey. \cite{2021ApJ...908..169G} identified 8 CT-AGNs by using the multiwavelength techniques, and found most of them had been misdiagnosed as low-luminosity AGNs. Therefore, in the CDFS survey, many missed CT-AGNs may still be hidden among the low-luminosity AGNs.
The motivation of our work is to find out  these hidden CT-AGNs by using a ML algorithm in the CDFS survey. 

The structure of the paper is as follows. In Section~\ref{sec:sample}, we described the data used in this work and constructed a reliable classification sample for training ML algorithms.
Section~\ref{sec:ML} detailed the input parameters of our ML algorithm, introduced the ML algorithm we used, trained the model in the training set, and evaluated it on the test set.
We have applied our well-trained algorithm to a subset of low-luminosity AGNs with the goal of finding out these hidden CT-AGNs in Section~\ref{sec:identifying}.
In Section~\ref{sec:discussion}, we discuss the fraction of CT-AGNs in the CDFS and the different properties of CT-AGNs and the non-CT-AGNs in host galaxies.
Finally, we present a brief summary of this work in Section~\ref{sec:summary}.
We adopt a concordance flat $\Lambda$-cosmology with ${\rm H_0 = 67.4 \ km\ s^{-1}\ Mpc^{-1}}$, $\Omega_{\rm m} = 0.315$, and $\Omega_\Lambda = 0.685$ \citep{2020A&A...641A...6P}.

\section{ Data and Sample} \label{sec:sample}

\begin{splitdeluxetable*}{cllcccccccccBrrcrcccccc}
	\tabletypesize{\scriptsize}
	\tablewidth{0pt} 
	\tablecaption{Physical Properties and Classification of Our AGN Sample.\label{Tab:summary}}
	\tablehead{
		\colhead{XID} & \colhead{R.A.} & \colhead{Decl.} & \colhead{z} & \colhead{z\_type}&\colhead{Counts}&\colhead{Counts\_soft}&\colhead{Counts\_hard}&\colhead{Time\_soft}&\colhead{Time\_hard}&\colhead{HR}&\colhead{$\Gamma$}&\colhead{$\log \mathrm{L}_{2-10}$}&\colhead{$\log \mathrm{L}_{6~\micron}$}&\colhead{$\log  (\mathrm{L}_{6~\micron}/\mathrm{L}_{2-10})$}&\colhead{$\log \mathrm{M_\star}$}&\colhead{$\log \mathrm{SFR}$}&\colhead{Class-trad}&\colhead{Quality}&\colhead{Class-ML}&\colhead{Probability}&\colhead{Pure CT-AGN}\\
		&\colhead{($^\circ$)}&\colhead{($^\circ$)}&&&&&&\colhead{(Ms)}&\colhead{(Ms)}&\colhead{}&&\colhead{($\mathrm{erg\ s^{-1}}$)}&\colhead{($\mathrm{erg\ s^{-1}}$)} & \colhead{} &\colhead{($\mathrm{M}_\odot$)} & \colhead{($\mathrm{M}_\odot \ \mathrm{yr}^{-1}$)} &\colhead{}&\colhead{}&
	} 
	\colnumbers
	\startdata 
	45& 52.967626 & -27.696263 &  0.857 & phot &$201^{+20}_{-19}$&$102^{+14}_{-13}$& $99^{+16}_{-15}$&0.394&0.494&$-0.12^{+0.09}_{-0.09}$&$1.3^{+0.21}_{-0.20}$&43.20 & 43.77 & 0.57 & 10.65 & 0.92 & non-CT-AGN&Insecure&non-CT-AGN&0.101&no\\
    52& 52.974346 & -27.706971 & 3.62 & phot &$67^{+19}_{-18}$&$56^{+13}_{-12}$& 32&0.893&1.09& -0.36& 2.49&43.95 & 44.97& 1.03& 10.96 & 1.30& non-CT-AGN&Insecure&non-CT-AGN&0.055& no\\
    66& 52.982191 & -27.719719 & 0.62 & spec &$71^{+27}_{-26}$& 46&84&2.61&3.14&0.20&1.4& 41.59 & 42.94 & 1.35& 10.16 &0.80& non-CT-AGN&Insecure&CT-AGN&0.730& yes\\
     68& 52.983105 & -27.75681 & 2.328 & phot &81& 14&$58^{+25}_{-23}$&4.93&5.53&0.57&-1.37&41.94 &44.11 & 2.17 &10.29 & 1.47 & non-CT-AGN&Insecure&CT-AGN&0.876& yes\\
     69& 52.984622 & -27.783408 & 0.889 & phot &$123^{+27}_{-25}$&46&$96^{+23}_{-22}$&5.27&5.63&0.33&-0.17&42.02& 44.38& 2.36 & 10.56 & 0.70 & non-CT-AGN&Insecure&CT-AGN&0.857& yes\\
     71& 52.984961 & -27.735114 &  1.178 & phot &$278^{+31}_{-30}$&$126^{+18}_{-17}$&$152^{+26}_{-25}$&3.56&4.10&$0.02^{+0.11}_{-0.11}$& $0.96^{+0.23}_{-0.23}$&42.68 & 43.11&0.43 & 9.22 & 0.21 & non-CT-AGN&Insecure&non-CT-AGN&0.220&no\\
     72& 52.986873 & -27.850233 &  1.692 & phot &$276^{+31}_{-30}$&$131^{+19}_{-18}$&$145^{+26}_{-24}$&5.39&5.80&$0.01^{+0.11}_{-0.11}$&$0.95^{+0.22}_{-0.22}$& 42.80 & 42.81& 0.01 & 10.42 & -0.06 & non-CT-AGN&Insecure&non-CT-AGN&0.063& no\\
      73& 52.987747 & -27.852295 & 2.509 & phot &$879^{+42}_{-41}$&$213^{+21}_{-20}$&$667^{+37}_{-35}$& 5.37&5.78&$0.49^{+0.04}_{-0.04}$&$-0.07^{+0.11}_{-0.11}$& 43.41 & 45.25 & 1.84 & 10.29 & 0.88& non-CT-AGN&Secure&CT-AGN&0.422& \nodata\\
      74& 52.989399 & -27.704551 & 1.581 & phot &$142^{+34}_{-32}$&28&$136^{+30}_{-28}$&2.02&2.41&0.61&-1.18& 42.71 & 42.76& 0.06 & 10.93&0.89 & non-CT-AGN&Insecure&non-CT-AGN&0.099& no\\
       76& 52.990417 & -27.711208 & 0.656 & phot &$95^{+27}_{-25}$&$39^{+15}_{-13}$&$56^{+23}_{-22}$&2.64&3.12&$0.09^{+0.27}_{-0.26}$&$0.81^{+0.63}_{-0.64}$&41.86 & 41.92 & 0.06 & 9.64 &0.12& non-CT-AGN&Insecure&CT-AGN&0.577&no\\
       77& 52.990482 & -27.702572 & 0.665 & spec & $543^{+37}_{-35}$&$337^{+27}_{-25}$&$207^{+27}_{-26}$&1.95&2.32&$-0.32^{+0.07}_{-0.07}$&$1.67^{+0.15}_{-0.16}$& 42.62 &42.87& 0.25 & 10.78 &0.43 & non-CT-AGN&Secure&non-CT-AGN&0.063& \nodata\\
       78& 52.991502 & -27.792961 & 1.85& phot & $107^{+23}_{-22}$&$42^{+13}_{-12}$& $66^{+20}_{-19}$&5.61&5.84&$0.21^{+0.20}_{-0.20}$&$0.59^{+0.45}_{-0.42}$&42.75 & 44.20 & 1.45 & 10.88&0.85& non-CT-AGN&Insecure&non-CT-AGN&0.473& no\\
       81& 52.992819 & -27.844873 &3.309& phot &$3580^{+68}_{-67}$&$1991^{+49}_{-48}$&$1592^{+48}_{-47}$&5.46&5.74&$-0.14^{+0.02}_{-0.02}$&$1.28^{+0.04}_{-0.04}$&44.48 & 45.73&1.25& 11.62 &1.46& non-CT-AGN&Secure&non-CT-AGN&0.079& \nodata\\
       82& 52.993095 & -27.909871 &2.282& phot & $160^{+27}_{-26}$&$103^{+17}_{-16}$& $57^{+22}_{-21}$&2.23&2.58&$-0.35^{+0.18}_{-0.18}$&$1.72^{+0.46}_{-0.46}$&43.23 & 45.35& 2.11& 10.97 &2.03 &non-CT-AGN&Insecure&non-CT-AGN&0.332& no\\
       86& 52.997546 & -27.781034 &0.733& spec & $167^{+24}_{-23}$&$99^{+15}_{-14}$& $67^{+19}_{-18}$&5.48&5.68&$-0.21^{+0.15}_{-0.15}$&$1.43^{+0.32}_{-0.32}$&41.81 & 41.78 & -0.02 & 10.49 &-0.01& non-CT-AGN&Insecure&non-CT-AGN&0.471& no\\
       92& 53.002584 & -27.704355 &1.905& phot & $171^{+29}_{-28}$&$99^{+17}_{-16}$&$72^{+24}_{-23}$&2.82&3.24&$-0.22^{+0.18}_{-0.17}$&$1.46^{+0.41}_{-0.39}$&42.99 & 44.03 & 1.04 & 10.86 &1.48 & CT-AGN&Secure&non-CT-AGN&0.370& \nodata\\
       93& 53.002932 & -27.856037 &2.421& phot & $49^{+22}_{-21}$&34&60&5.50&5.74&0.26&1.4& 42.38 & 44.95 & 2.57& 10.87 &1.12&non-CT-AGN&Insecure&CT-AGN&0.853& yes\\
        133& 53.020614 & -27.742022 &3.472& spec & $71^{+22}_{-21}$&20&$66^{+19}_{-18}$&5.60&5.82&0.52&-0.88& 42.17& 44.52 & 2.35 & 9.72&1.35 &CT-AGN&Secure&CT-AGN&0.859& \nodata\\
         137& 53.022557 & -27.792694 &2.636& spec &$38^{+14}_{-13}$&$28^{+9}_{-8}$&27&6.01&6.02&-0.02&1.7& 42.34 & 42.35 &0.01 & 10.70 &0.64 &non-CT-AGN&Insecure&non-CT-AGN&0.054& no\\
 $\vdots$&$\vdots$&$\vdots$&$\vdots$&$\vdots$&$\vdots$&$\vdots$&$\vdots$&$\vdots$&$\vdots$&$\vdots$&$\vdots$&$\vdots$&$\vdots$&$\vdots$&$\vdots$&$\vdots$&$\vdots$&$\vdots$&$\vdots$&$\vdots$&$\vdots$\\
	\enddata
	\tablecomments{This table is available in its entirety in the machine-readable format. Only a portion of this table is shown here to demonstrate its form and content. Column (1) is X-ray IDs from the 7~Ms catalog \citep{2017ApJS..228....2L}. Columns (2) and (3) contain the R.A. and decl. of the X-ray sources, respectively.  Column (4) contains the redshift that we used. Column (5) specifies the type of redshift, where “spec" indicates spectral redshift and “photo" represents photometric redshift. Columns (6), (7) and (8) show the net counts and errors for the X-ray source in the 0.5--7 keV band, the soft net counts and errors in the 0.5--2 keV band, and the hard net counts and errors in the 2--7 keV band, respectively. The absence of errors in the net counts is because these sources were undetected in the given band. These values indicate the 90\% confidence upper limits on the net counts. Columns (9) and (10) show the effective exposure times for the soft (0.5--2 keV) and hard (2--7 keV) bands, respectively. Column (11) presents the HRs and their errors. Column (12) is the photon index provided by the 7~Ms catalog \citep{2017ApJS..228....2L}. Columns (13) and (14) show the logarithm of 2-10 keV and 6~\micron~luminosity for AGNs in the rest-frame. Column (15) represents the MIR-X-ray luminosity ratio. Columns (16) and (17) contain the logarithm of the stellar mass and SFR for the host galaxy of the AGN, respectively. Columns (18) and (19) are the classification and classification quality that were reliably diagnosed by traditional methods. Among them, “Secure” means data in training and test sets for reliable classification, while “Insecure” refers to data for application prediction. Column (20) shows the classification results of our ML algorithm, which fall into two categories: CT-AGN and non-CT-AGN. Column (21) represents the probability of being classified as CT-AGN in our well-trained ML algorithm. Column (22) indicates pure CT-AGN, if the probability of CT-AGN exceeds 0.664 in Low-Luminosity sample, mark yes; otherwise, mark no; “\nodata" denotes the sources present in both the training and test sets.}
\end{splitdeluxetable*}

\subsection{X-ray data}
Due to the definition of CT-AGNs being in the X-ray band, identifying them often requires the use of their X-ray data. In this work, the X-ray data for AGNs is mainly provided by the 7~Ms catalog of \cite{2017ApJS..228....2L}, including full band (0.5--7~keV) Net counts, soft band (0.5--2~keV) Net counts, hard band (2--7~keV) Net counts, column density, photon index ($\Gamma$), apparent 0.5–7 keV luminosity and so on. 
In this work, the X-ray data used are all listed in Table~\ref{Tab:summary}.
Moreover, we also use the column density of some sources provided by \cite{2017ApJS..232....8L}, \cite{2019A&A...629A.133C}, \cite{2019ApJ...877....5L}, and \cite{2021ApJ...908..169G}. 
In this work, we need the observed luminosities of AGNs in the 2--10~keV band, which is not directly provided by the 7~Ms catalog \citep{2017ApJS..228....2L}. Fortunately, it can be indirectly calculated from apparent 0.5--7 keV luminosities, with the conversion formula being as follows:
\begin{eqnarray*}
	{\rm L}_{2-10} =& {\rm L}_{0.5-7} \frac{{\ln 10 - \ln 2}}{{\ln 7 - \ln 0.5}}, \Gamma = 2 \\
	{\rm L}_{2-10} =& {\rm L}_{0.5-7} \frac{{10^{-\Gamma+2} - 2^{-\Gamma+2}}}{{7^{-\Gamma+2} - 0.5^{-\Gamma+2}}}, \Gamma \neq 2.
\end{eqnarray*}
The 2--10 keV luminosities are listed in the column 13 of Table~\ref{Tab:summary}.

\subsection{MIR data}
AGNs often exhibit bright MIR emissions primarily because of the thermal radiation from hot dust in the torus. The MIR band has a lower optical depth compared to other wavelengths, such as soft X-rays, so the emission from AGNs in this band is not as strongly absorbed or suppressed. This makes it a good tracer for the intrinsic X-ray emission, as both are related to the radiation from the accretion disk. Therefore, MIR data is essential for training the ML algorithm to identify CT-AGN. The MIR data used in this work, specifically the 6~$\micron$ luminosities of AGNs obtained through Spectral Energy Distribution (SED) decomposition, is provided by \cite{2020MNRAS.492.1887G}. The 6~$\micron$ luminosities of AGNs are listed in the column 14 of Table~\ref{Tab:summary}.

\subsection{Redshift selection and Luminosity correction}
Redshifts are a fundamental parameter in the study of galaxies and AGNs, being closely related to the luminosity distance, and thus using accurate and reliable redshifts is imperative for effectively training the ML algorithm.
In this work, we use the latest redshifts, which are provided by \cite{2020MNRAS.492.1887G}. The redshifts and types of redshifts are listed in the columns 4 and 5 of Table~\ref{Tab:summary}.

Since we have used more accurate redshifts relative to the 7~Ms catalog, the observed X-ray luminosities need to be corrected. The newly corrected X-ray luminosity is derived by
\begin{equation*}
	\frac{L_{\rm{new}}}{L_{\rm{old}}} = \frac{R(z)^2(1+z)^{\Gamma-2}}{R(z_{\rm{7Ms}})^2(1+z_{\rm{7Ms}})^{\Gamma-2}},
\end{equation*}
where $L_{\rm{new}}$ and $L_{\rm{old}}$ are newly corrected X-ray luminosity and observed X-ray luminosity derived by 7 Ms catalog, $z$ and $z_{\rm{7Ms}}$ are our redshift and redshift from 7 Ms catalog, and $R(z)$ is the luminosity distance.


\subsection{The AGN sample}

In the 7~Ms catalog, 711 sources were diagnosed as X-ray AGNs.
However, some so-called normal galaxies in the 7Ms catalog may be hidden AGNs. \cite{2018ApJ...868...88D} identified 12 AGNs through X-ray variability, and \cite{2020MNRAS.492.1887G} also found 6 AGNs using SED fitting. So there are currently a total of 728 AGNs in the CDFS. Only 464 AGNs include the MIR data we used in this work. 
62 sources have been classified as CT-AGNs among the 464 AGNs based on their column densities or previous studies \citep{2017ApJS..232....8L,2019ApJ...877....5L,2019A&A...629A.133C,2021ApJ...908..169G}.


To effectively train a ML algorithm for classifying AGNs, we need to compile a sample with secure and accurate classifications.
Therefore, we select the AGN subset with secure classifications based on one of the following 3 criteria:
\begin{enumerate}[label = (\arabic*),itemjoin=\\]
	\item Sources in the AGNs sample of \cite{2021ApJ...908..169G};
	\item Sources identified as CT-AGNs by previous work;
	\item Sources with X-ray Net counts (0.5--7~keV) greater than 300.
\end{enumerate}
The first criterion is used for the sample from \cite{2021ApJ...908..169G}, whose sample offers a secure classification for both CT-AGNs and Compton-thin AGNs. The second criterion involves the selection of CT-AGNs that have been identified as reliable through previous X-ray diagnostic studies.
In X-ray spectral fitting analysis, the quality of the spectrum is directly related to the reliability of the column density, which in turn affects subsequent classification. Specifically, the more the net counts of the source, the better the statistical significance of the obtained X-ray spectrum. This makes the classifications more secure. Therefore, we have implemented the third selection criterion.

Based on our three criteria, a total of 210 AGNs are securely classified, including 62 CT-AGNs. This securely classified sample of AGNs is used to train our ML algorithm.
Additionally, 254 AGNs do not meet any of the established criteria. The classification and classification quality of the AGNs in our sample are presented in columns 18 and 19 of Table~\ref{Tab:summary}.
Given their low photon counts, many of them are likely to be low X-ray luminosity AGNs. This implies that a considerable number of CT-AGNs may be hidden within this group. We will consider these 254 AGNs as a subset of low-luminosity AGNs, in Section~\ref{sec:Application}, we will use our well-trained ML algorithm to search for CT-AGNs within this group.

\section{ML algorithm}\label{sec:ML}
This work aims to use an ML algorithm to identify CT-AGNs, which is inherently a supervised classification task within the field of ML. Considering our training dataset comprises a small sample size of only 210 sources, it is crucial that the chosen algorithm can deliver robust performance even with limited data and is capable of handling high-dimensional data effectively. Among the various algorithms that can satisfy these requisites, the Random Forest \citep{2001MachL..45....5B}algorithm stands out as an optimal choice.

In this section, we begin by detailing the input parameters used in our ML algorithm. Next, we also introduce the Random Forest algorithm. Following that, we trained the model in training set. 
Finally, we will apply the well-trained to the test set for evaluation.

\subsection{Input parameter used}
To achieve the best classification results with the ML algorithm, the input parameters used are usually closely related to the classification results. In this work, column density is a key factor for classification, therefore, we selecte a set of physical quantities that are relevant to column density as input parameters for our ML algorithm. 
Here's how each parameter relates to the column densities of AGNs:
\begin{itemize}
    %
	\item \textbf{Hardness Ratio (HR)}: The HR a metric used to quantify the relative intensity of hard X-rays compared to soft X-rays. It is defined as (H -- S)/(H + S), H represents the count rate\footnote{Count rates = Counts/Time.} in the hard X-ray band (2--7 keV), and S represents the count rate in the soft X-ray band (0.5--2 keV). 
    HR reflects the X-ray spectral characteristics of AGN and can be used to estimate the absorbing column density. A higher HR may indicate significant gas or dust obscuration along the line of sight, implying a higher column density. Conversely, a lower HR suggests a lower column density. HRs and their errors\footnote{
    The error in HR is approximately:
\(
E_{\text{HR}} = \frac{2 \sqrt{S^2 \cdot E_H^2 + H^2 \cdot e_S^2}}{(H + S)^2}
\),
\(
e_{\text{HR}} = \frac{2 \sqrt{S^2 \cdot e_H^2 + H^2 \cdot E_S^2}}{(H + S)^2}
\).
Here, \(E_{\text{HR}}\) denotes the upper error bound of the hardness ratio (HR), and \(e_{\text{HR}}\) denotes the lower error bound. The quantities \(E_{\text{H}} \) and \( e_{\text{H}}\) denote the upper and lower uncertainties of the hard count rates, while \(E_{\text{S}}\) and \( e_{\text{S}} \) represent the upper and lower uncertainties of the soft count rates.} are listed in column 11 of Table~\ref{Tab:summary}. 

	\item \textbf{Redshift}: The Photon Index and HR both vary as a function of redshift \citep[e.g.,][]{2019ApJ...877....5L}. Due to the cosmic expansion, the X-ray emission from an AGN experiences redshift, which in turn affects the observed characteristics of the emission.
	\item \textbf{The luminosity of AGN in the MIR}: The MIR emission, which is less affected by dust and gas absorption, can provide a clear view of the AGN's intrinsic luminosity. In particular, the luminosity at 6~\micron~can clearly reflect the intrinsic luminosity of the AGN.
	\item \textbf{MIR–X-ray Luminosity Ratio}: Compared to X-ray emission, MIR emission is less affected by dust and gas absorption. Considering the correlation between X-ray and MIR emissions in AGNs, the MIR-X-ray luminosity ratio can effectively reflect the absorption of the gas. These ratios are listed in column 15 of Table~\ref{Tab:summary}.
\end{itemize}

\begin{figure*}
	\centering
	\includegraphics[width=\textwidth]{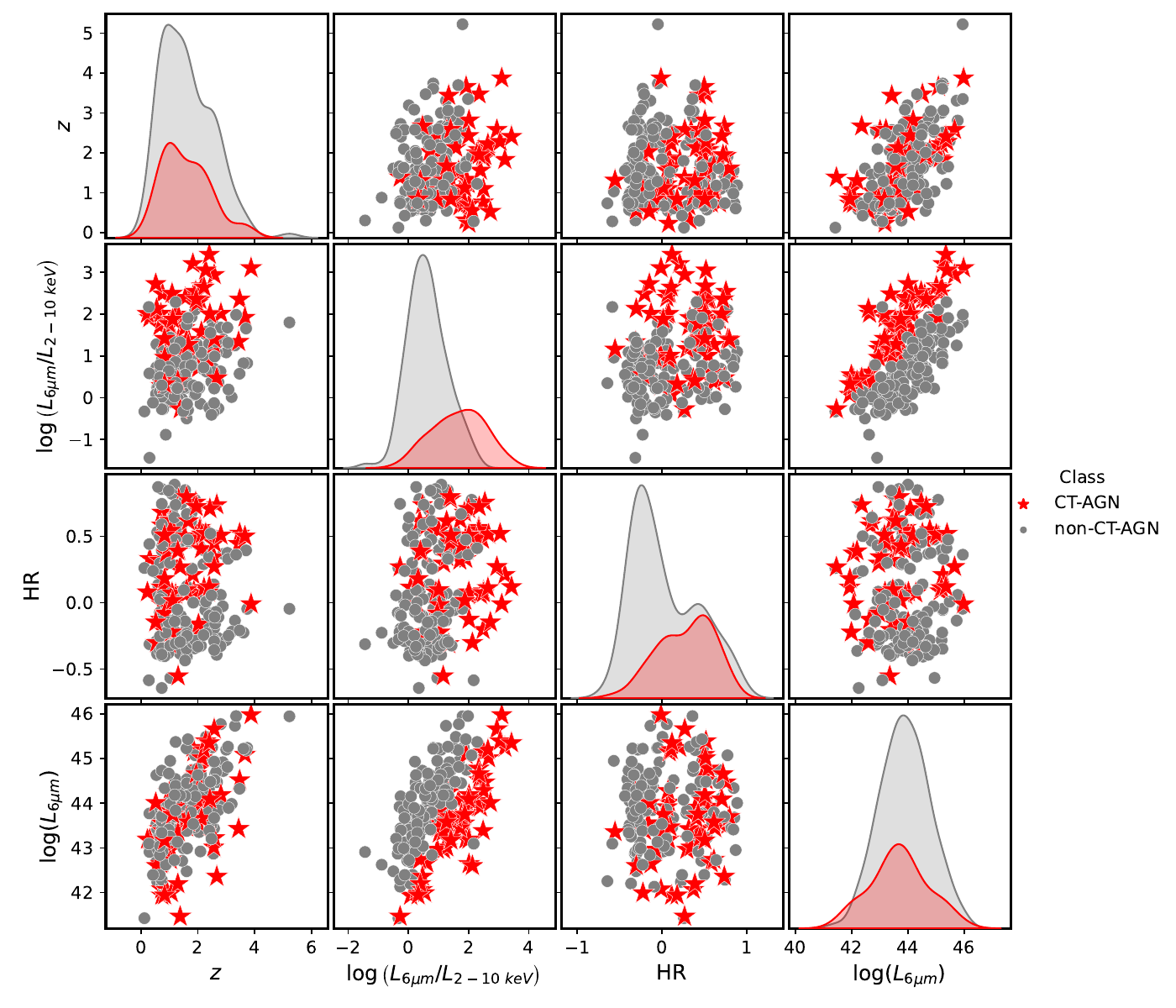}
	\caption{The distribution of input parameters for CT-AGNs and non-CT-AGNs in the training dataset. The red stars indicate CT-AGNs, and the gray circles represent non-CT-AGNs.}
   
	\label{fig:parameter}
\end{figure*}

In summary, we selected 4 physical quantities as the input parameters. Figure~\ref{fig:parameter} illustrates the distribution of CT-AGNs and Compton-thin AGNs in the input parameter space. 

\subsection{Random Forest}\label{sec:Random Forest}
\begin{figure*}
	\centering
	\includegraphics[width=0.95\linewidth]{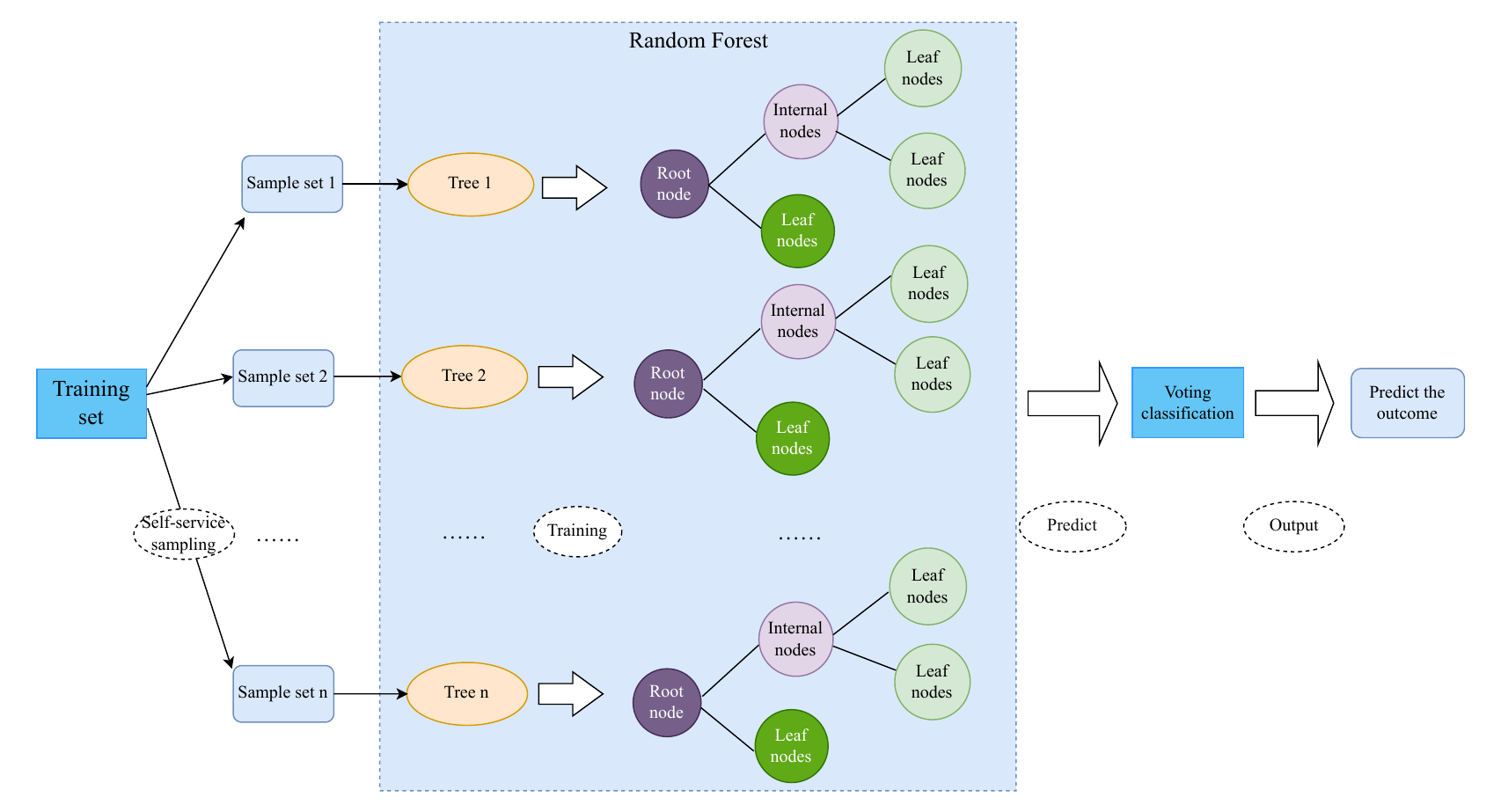}
	\caption{The Random Forest algorithm, with its principle shown in a cartoon diagram. 
    }
	\label{fig:Random-forest}
\end{figure*}

Random Forest, an ensemble learning classifier, is publicly accessible within the \textit{scikit-learn} library—an open-source \textit{Python} ML platform developed by \cite{2011JMLR...12.2825P}. This classifier combines multiple decision trees to improve classification performance. 
Each decision tree is constructed using a random subset of the training data, contributing its vote towards the class labels. The ensemble classifier's final prediction is determined by the majority vote across all trees, effectively synthesizing the collective predictions into a single, more accurate and robust outcome. 
Figure~\ref{fig:Random-forest} shows an overview of the Random Forest classifier.

A single decision tree classifier forms the backbone of Random Forests. The general idea of a single tree classifier is to find a set of optimal rules for partitioning the feature space to differentiate data points belonging to various classes. The working mechanism of a single decision tree involves applying new regulations that cause the dataset to split into two branches, creating a new node each time. As this process is recursive, the decision graph resembles an upside-down tree. Within this process, \textit{Gini impurity} is a key metric used for node splitting. The calculation formula of \textit{Gini impurity} is as follows:
\begin{equation*}
	\text{Gini}(D) = 1 - \sum_{i=1}^{n} p_i^2
\end{equation*}
where $D$ represents the dataset, $p_i$ is the probability of class $i$. 

During training, the decision tree calculates the \textit{Gini impurity} of each candidate segmentation point for each split node and selects the segmentation point that can minimize the \textit{Gini impurity} for splitting.
In addition to using \textit{Gini impurity} as the splitting criterion, we have also set other parameters: the random seed is used to ensure the experiment's reproducibility. By limiting the number of features considered at each split, we increase the randomness and generalization ability of the model. We control the growth of the tree by limiting the maximum depth and the number of leaf nodes to prevent the tree from becoming overly complex. With these adjustments, we aim to achieve the best model performance.


To effectively train our algorithm, we meticulously divided the dataset. In the AGN sample with secure classification, we randomly selected $70.0\%$ of the data for the training set, while the remaining $30.0\%$ was designated for the test set. Given our limited sample size, this division may result in some distribution bias between the training and test sets in terms of input parameters. To ensure that the training and test sets are drawn from the same distribution, we conducted the Kolmogorov-Smirnov test \citep[KS-test,][]{karson1968handbook} on each parameter utilized in both sets. All p-values associated with the parameters exceed the threshold of 0.05, indicating that these parameters from the training and test sets are indeed from the same distribution at a $95\%$ confidence level. This implies that our dataset partitioning is justified, and the training set can effectively serve as a valid representation of the test set.

\subsection{Model Training and evaluation}
To address the risk of overfitting inherent to small-sample datasets, we implemented a series of optimization strategies to refine model performance. By carefully fine-tuning model complexity, we found that setting the \textit{n\_estimators} parameter to 25 enabled the model to achieve an optimal balance: it yielded relatively high accuracy while preserving robust generalization capabilities. This approach successfully prioritized predictive accuracy while maintaining an appropriate trade-off between model performance and computational efficiency.
In this work, after repeated optimization, the parameters of the optimal model have been listed in Table~\ref{tab:Parameter}. The out-of-bag (OOB) score of the optimal model is 0.87, suggesting that the model exhibits a high level of accuracy when predicting the OOB samples, indicating robust generalization capabilities.
\begin{deluxetable}{lc}
	\label{tab:Parameter}
	\tablecaption{The optimal model parameter settings.}
	\tablewidth{0pt}
	\tablehead{
		\colhead{Parameter} & \colhead{Setting value} 
	}
	\decimalcolnumbers
	\startdata
    \textit{max\underline{~}features} & sqrt \\
    \textit{Gini impurity} & Gini \\
	\textit{bootstrap}&True \\  
	\textit{n\underline{~}estimators} & 25 \\
	\textit{max\underline{~}depth} & 6 \\
	\textit{min\underline{~}samples\underline{~}leaf} & 6 \\
	\textit{min\underline{~}samples\underline{~}split} & 5 \\
	\textit{max\underline{~}leaf\underline{~}nodes} & 8 \\
	\enddata
\end{deluxetable}
\subsection{ Classification Performance Evaluation}
To comprehensively evaluate the performance of our algorithm on the test set, we used three key statistical metrics: Accuracy, Precision, and Recall. The definitions of these metrics are as follows:
\begin{equation*}
\begin{aligned}
	&\text{Accuracy}  = \frac{\text{TP} + \text{TN}}{\text{TP} + \text{FP} + \text{TN} + \text{FN}},\\
	&\text{Precision}  = \frac{\text{TP}}{\text{TP} + \text{FP}},\\
	&\text{Recall} = \frac{\text{TP}}{\text{TP} + \text{FN}} ,
\end{aligned}
\end{equation*}
Where TP (True Positives) indicates the number of true positives, TN (True Negatives) indicates the number of true negatives, FP (False Positives) indicates the number of false positives, and FN (False Negatives) indicates the number of false negatives. The positive condition is defined as CT-AGN, and the negative condition is defined as non-CT-AGN. 
The higher the accuracy, the more precise our algorithm is in their classification. The higher the precision, the fewer CT-AGNs are misdiagnosed when using our algorithm for classification. The higher the recall rate, the fewer CT-AGNs are missed.
\begin{figure}
	\centering
	\includegraphics[width=1.0\linewidth]{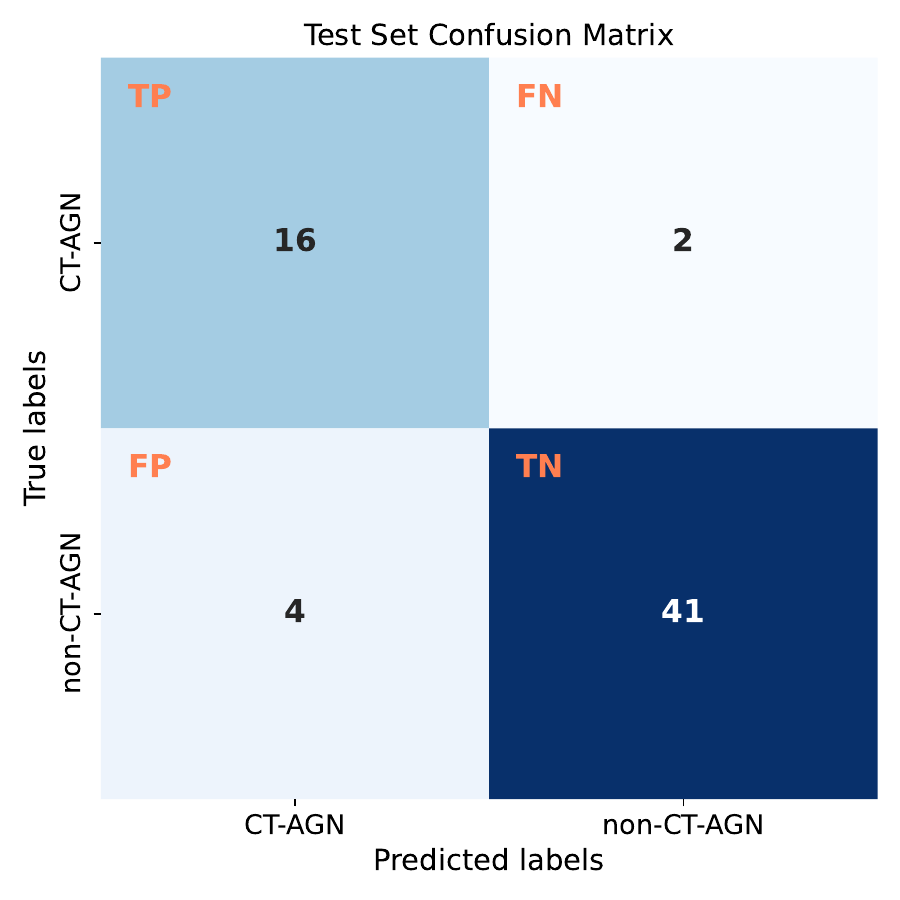}
	\caption{The confusion matrix computed from this test set shows the overall counts. The test set contains 63 AGNs.}
	\label{fig:Confusion-Matrix}
\end{figure}

We deploy our well-trained algorithm onto the test set and conducted a comparison with the actual classifications to assess the model's capacity for generalization and predictive accuracy. 
In Figure~\ref{fig:Confusion-Matrix}, we report the confusion matrix values accounting for the TN, FP, FN, and TP in the test set.
The accuracy of our algorithm on the test set is 0.90.
The precision and recall of our algorithm for predicting CT-AGNs on the test set are 0.80 and 0.89, respectively.
These results indicate that our algorithm performs well on the test set. 

\section{Identifying CT-AGNs With ML algorithm}\label{sec:identifying}

\subsection{Identification criteria for CT-AGNs}
\begin{figure}
	\centering
	\includegraphics[width=1.\linewidth,trim=10 0 10 0]{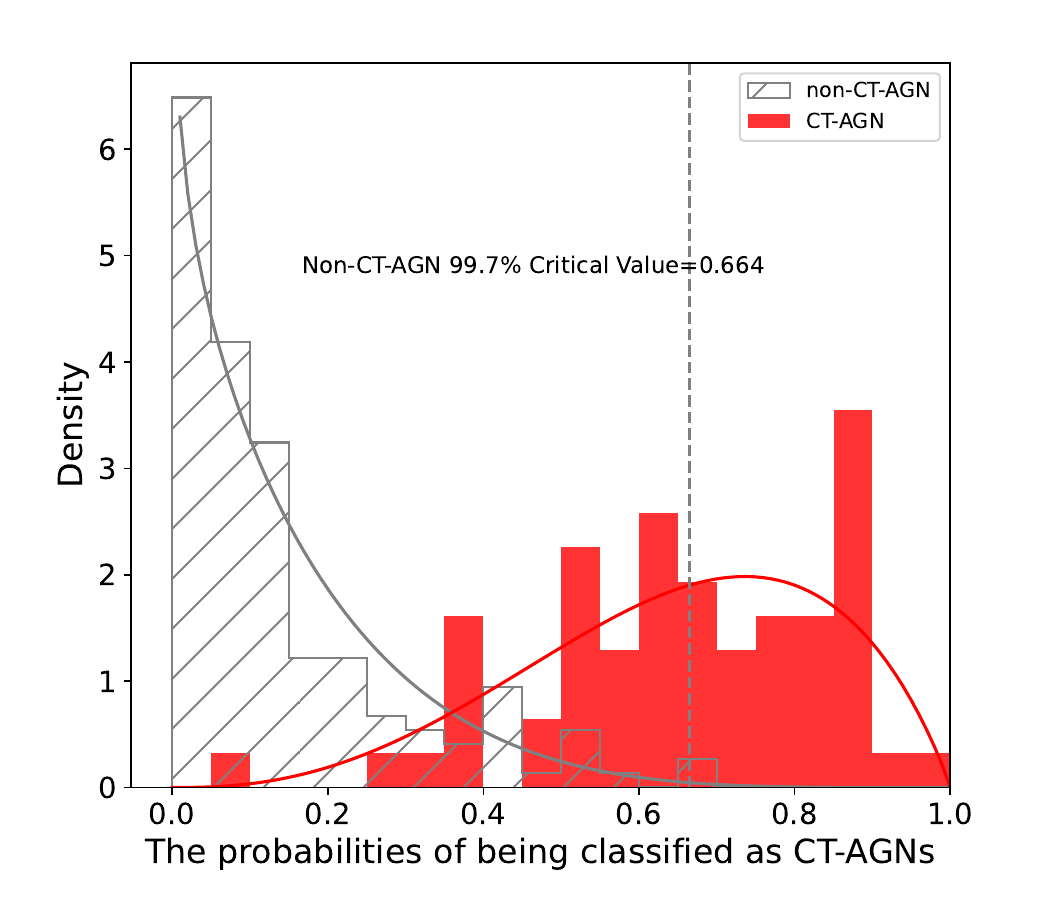}
	\caption{The distribution of the probability of being classified as CT-AGNs for different types of AGNs. The gray step histogram represents non-CT-AGNs, while the red solid histogram represents CT-AGNs. The gray vertical dashed line marks the critical value at the 99.7\% level on the right side of the non-CT-AGN distribution.}
	\label{fig:probability}
\end{figure}

We aim to identify and select a genuinely pure sample of CT-AGNs. However, using our well-trained algorithm, there were 4 sources misdiagnosed as CT-AGNs in the test set. To eliminate these FP, we used the \textit{prediction\underline{~}probe} function to determine the probabilities of being classified as CT-AGNs for securely classified sources. 
Figure~\ref{fig:probability} illustrates the distribution of the probability of being classified as CT-AGNs for different types of AGNs. The red solid histogram represents CT-AGNs, while the gray step histogram represents non-CT-AGNs.
It is evident that there is an overlapping region between CT-AGNs and non-CT-AGNs in the probability of being classified as CT-AGNs. When selecting the classification threshold in the overlapping region, some sources are typically misdiagnosed. 
Therefore, to ensure that as many CT-AGNs as possible are diagnosed while minimizing misdiagnoses, we needed to determine an appropriate threshold. 

Considering that the probability distribution output by the Random Forest algorithm typically conforms to a Beta distribution, we used the Beta function to fit the probability distributions for both CT-AGNs and non-CT-AGNs. Ultimately, we selected the critical value at the 99.7\% on the right side of the Beta distribution followed by non-CT-AGNs as the threshold for identifying CT-AGNs, which was 0.664. This implies that any source with a probability greater than or equal to 0.664 is classified as a CT-AGN. This approach helps to minimize FP while maximizing the identification of pure CT-AGNs.

\subsection{Applied to Low-Luminosity Sample}\label{sec:Application}

We applied our well-trained algorithm on the 254 Low-luminosity AGNs and estimated the probability that each AGN could be a CT-AGN using the \textit{prediction\underline{~}probe} function. The corresponding diagnostic outcomes are detailed in columns 20 and 21 of Table~\ref{Tab:summary}.

Within the subset of low-luminosity AGNs, 126 AGNs have a probability greater than 50\% of being classified as CT-AGNs. However, it is possible that some sources classified as CT-AGNs may have been misdiagnosed. This implies that the sample composed of these 126 sources might be an impure CT-AGN sample. To select a pure sample of CT-AGNs, we applied the probability threshold previously established for pure CT-AGNs. Based on this criterion, we found that 67 sources had probabilities exceeding the threshold, thus they are identified as CT-AGNs. They are listed in column 22 of Table~\ref{Tab:summary}.
Among the 67 CT-AGNs identified by our ML algorithm, most have not been already known CT-AGNs. However, the three sources in the 7~Ms catalog \citep[namely XID 68, 414, and 575;][]{2017ApJS..228....2L} have been identified as CT-AGN, with $N_\mathrm{H} > 10^{24}\ cm^{-2}$. Therefore, we have identified 64 hitherto unknown CT-AGNs.


\section{Discussion} \label{sec:discussion}
\subsection{Feature importance scores}
\begin{figure}
	\centering
	\includegraphics[width=1.\linewidth,trim=40 0 40 0]{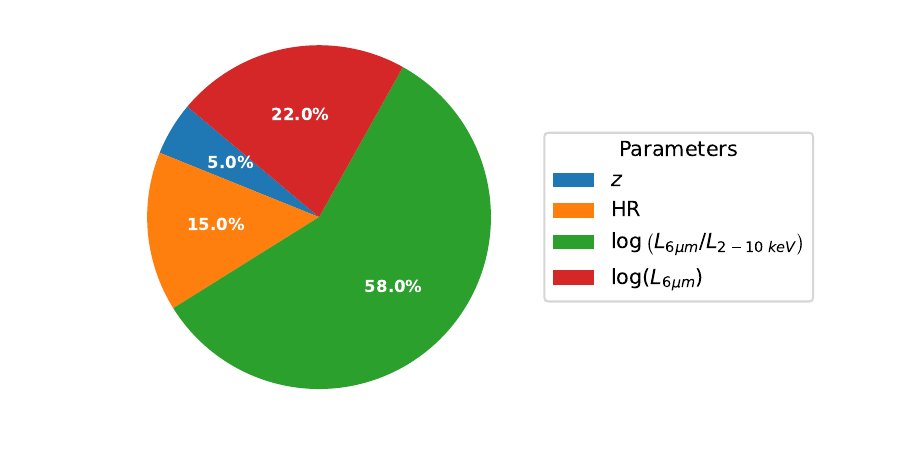}
	\caption{The display of the importance of all parameters. The MIR-X-ray Luminosity Ratio is the preeminent factor contributing to the algorithm's predictive capacity, with the 6~\micron~luminosity closely trailing behind.}
	\label{fig:importance}
\end{figure}

The input parameters we selected are physical quantities that are all related to column density, playing a crucial role in identifying CT-AGNs. For instance, the column density can be roughly estimated using the HR and redshift assuming an intrinsic AGN X-ray spectrum \cite[e.g.,][]{2019ApJ...877....5L,2022ApJ...941...97V}. We used feature importance scores to quantify the contribution of these input parameters to the model's classification capability, as depicted in Figure~\ref{fig:importance}.
Among the parameters considered, the MIR-X-ray luminosity ratio stands out as the most significant contributor to the model's classification performance, accounting for 58.0\% of the importance score. 
In other words, in our classifier model, the MIR-X-ray luminosity ratio is the most important input parameter for distinguishing between CT-AGNs and Compton-thin AGNs.
The importance scores for the other parameters are as follows:  The 6~\micron~luminosity has a score of 22.0\%, HR has a score of 15.0\%, and redshift $z$ has a score of 5.0\%.

\subsection{The fraction of CT-AGNs}

In the CDFS, traditional methods identified 79 CT-AGNs \citep{2017ApJS..232....8L,2019ApJ...877....5L,2019A&A...629A.133C,2021ApJ...908..169G}, which constitute about 11\%(79/728) of the total AGN population. Within this field, we selected a sample of 464 AGNs, among which 62 were identified as CT-AGNs by traditional methods. By using our ML algorithm, we successfully identified 67 pure CT-AGNs, which includes three already known CT-AGNs. This increases the fraction of CT-AGNs in our sample to 27.8\%, aligning closely with the theoretical expectation of CXB. However, in the CDFS field, the number of CT-AGNs amounts to 146, accounting for about 20.0\% of the total AGNs. 
Although this work has significantly increased the fraction of CT-AGNs in this field, it still falls short of reaching the theoretical expections of the CXB. Therefore, further research work may be needed in the future to increase this fraction.

\subsection{The Properties of CT-AGN Host Galaxies}

CT-AGNs are considered to represent an early phase of the evolutionary scenario for AGNs \citep[e.g.,][]{2023PASP..135a4102G}. Many studies have shown that the evolution of AGNs is intricately linked to their surrounding environment, suggesting a coevolution scheme between host galaxies and AGNs \citep[e.g.,][]{2013ARA&A..51..511K}, such as M$_\mathrm{BH}$--$\sigma_\star$ \citep[e.g.,][]{2000ApJ...539L...9F},  AGN feedback \citep[e.g.,][]{2006MNRAS.370..645B}. 
In addition, some studies have indicated that CT-AGNs tend to be hosted in galaxies with a significant amount of dense gas \citep[e.g.,][]{2015ApJ...814..104K,2017MNRAS.468.1273R}. The radiation from AGNs can compress the surrounding gas, potentially triggering star formation activity within their host galaxies. \cite{2006ApJS..163....1H} has suggested that the host galaxies of CT-AGNs appear to be at intense star formation. \cite{2012ApJ...755....5G} demonstrated that the host galaxies of CT-AGNs in the local universe exhibit a high level of star formation. However, \cite{2022A&A...666A..17G} suggests that the host galaxies of high-redshift AGNs are different from those of AGNs in the local universe. In the CDFS, \cite{2021ApJ...908..169G} indicates similar distributions in terms of stellar masses and SFRs between the host galaxies of CT-AGNs and non-CT-AGNs in their sample of 51 AGNs. Despite the findings, the sample size of CT-AGNs is too small to ensure that the conclusions drawn are statistically robust. Therefore, we plan to use a larger sample size than that used in \cite{2021ApJ...908..169G} to conduct a comparison of the SFRs and stellar masses of host galaxies between CT-AGNs and non-CT-AGNs.

\begin{figure}
	\centering
	\includegraphics[width=1.0\linewidth]{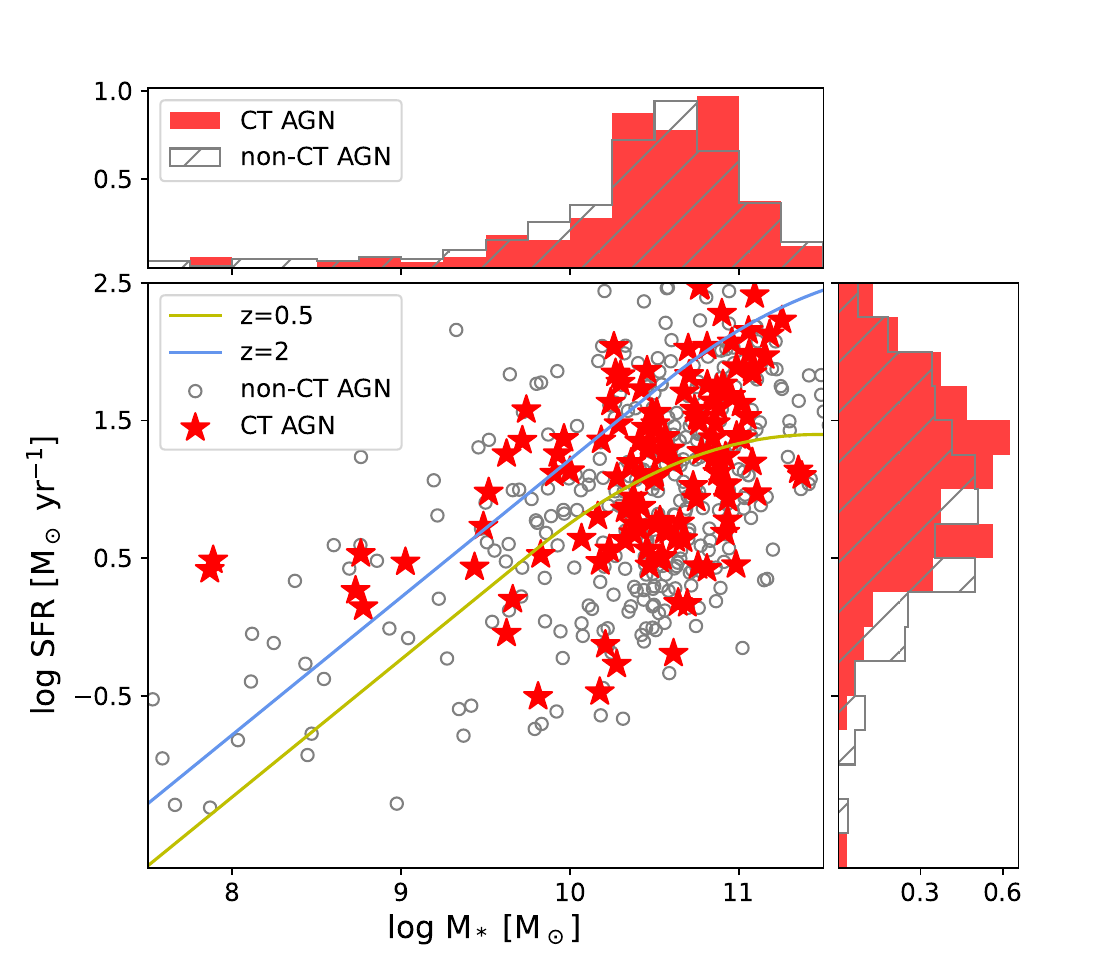}
	\caption{The relationship of the main sequence in AGN host galaxies. The solid red stars indicate the CT-AGNs, and the open gray circles represent non-CT-AGNs.}
	\label{fig:SFRvsMass}
\end{figure}
In this comparison, the sample comprises a total of 464 AGNs. The CT-AGNs within this sample include 67  pure CT-AGNs and 62 known CT-AGNs. Others are regarded as non-CT-AGNs. Figure~\ref{fig:SFRvsMass} presents relationships, known as main sequences, between SFRs and stellar masses of the host galaxies for both CT-AGNs and non-CT-AGNs. We used the KS-test to examine the distributions of stellar masses for their host galaxies, obtaining a $p$-value of 0.506. This result suggests that there is a same distribution of stellar masses among the host galaxies of different AGNs. Subsequently, we repeated the KS-test to examine their SFR distributions, yielding a $p$-value of  $5.209\times10^{-3}$. This indicates that there is significant discrepancy in the SFR distributions across the host galaxies of different AGNs. The average and median $\log$ SFRs of the host galaxies of CT-AGNs are 1.13 and 1.19, respectively. For non-CT-AGNs, the average and median SFRs of their host galaxies are 0.88 and 0.95, respectively. These results indicate that the star-forming activity in the host galaxies of CT-AGNs may be more intense compared to those of non-CT-AGNs. Our findings are consistent with the claim made by \cite{2012ApJ...755....5G}, indicating a possible correlation between the intense star formation and the presence of CT-AGNs. The possible reason is that the radiation from CT-AGNs compresses the surrounding dense gas, triggering star formation activity in their host galaxies.
\section{Summary}\label{sec:summary}
In this work, we used Random Forest algorithm to identify 67 CT-AGNs in the CDFS field, significantly increasing the fraction of CT-AGNs and aligning it with the theoretical predictions of the CXB.

Firstly, in the CDFS field, we constructed a sample of AGNs with multiwavelength data. We divided the sample into subsets with secure and insecure classifications. We trained and tested our Random Forest algorithm on the subset that could be securely classified as CT-AGNs and non-CT-AGNs. Our algorithm achieved an accuracy rate of 90\% on the test set after training. These results indicate that our algorithm performs well. Subsequently, we applied our algorithm to the subset of low-luminosity AGNs with insecure classifications and successfully identified 67 new CT-AGNs. This result significantly increased the fraction of CT-AGNs in the CDFS. Finally, we compared the properties of host galaxies between CT-AGNs and non-CT-AGNs in our sample and found that the host galaxies of CT-AGNs exhibit higher levels of star formation activity than those of non-CT-AGNs.

 \section*{Acknowledgments}
 We sincerely thank the anonymous reviewer for useful suggestions.
 We acknowledge the support of \emph{National Nature Science Foundation of China} (No.  12303017).
 This work is also supported by \emph{Anhui Provincial Natural Science Foundation} project number 2308085QA33.
 QSGU is supported by the \emph{National Natural Science Foundation of China} (12121003, 12192220, and 12192222).
 Yongyun Chen is grateful for financial support from the \emph{National Natural Science Foundation of China} (No. 12203028). 
 Yongyun Chen is grateful for funding for the training Program for talents in Xingdian, Yunnan Province (2081450001).
 HCF acknowledges support from \emph{National Natural Science Foundation of China} (12203096)
 Rui Li acknowledges the \emph{National Nature Science Foundation of China} (No. 12203050).
 H.T.W is supported by the \emph{Hebei Natural Science Foundation of China} (Grant No. A2022408002).


%

\vspace{5mm}


\software{astropy \citep{2013A&A...558A..33A,2018AJ....156..123A},  
           numpy, pandas,  matplotlib \citep{2007CSE.....9...90H}
          }



%


\bibliography{bibtex}{}
\bibliographystyle{aasjournal}








\end{document}